# An Activity Recognition Framework for Continuous Monitoring of Non-Steady-State Locomotion of Individuals with Parkinson's Disease

Mahdieh Kazemimoghadam[1] and Nicholas P. Fey[2*]

[1] Department of Bioengineering at The University of Texas at Dallas, Richardson, TX 75080, USA; mahdieh.kazemimoghadam@utdallas.edu.

[2*] Department of Mechanical Engineering, The University of Texas at Austin, Austin, TX 78712, USA; email: nfey@utexas.edu

*Abstract*— Fundamental knowledge in activity recognition of individuals with motor disorders such as Parkinson's disease (PD) has been primarily limited to detection of steady-state/static tasks (e.g., sitting, standing, walking). To date, identification of non-steady-state locomotion on uneven terrains (stairs, ramps) has not received much attention. Furthermore, previous research has mainly relied on data from a large number of body locations which could adversely affect user convenience and system performance. Here, individuals with mild stages of PD and healthy subjects performed non-steady-state circuit trials comprising stairs, ramp, and changes of direction. An offline analysis using a linear discriminant analysis (LDA) classifier and a Long-Short Term Memory (LSTM) neural network was performed for task recognition. The performance of accelerographic and gyroscopic information from varied lower/upper-body segments were tested across a set of user-independent and user-dependent training paradigms. Comparing the F1 score of a given signal across classifiers showed improved performance using LSTM compared to LDA. Using LSTM, even a subset of information (e.g., feet data) in subject-independent training appeared to provide F1 score > 0.8. However, employing LDA was shown to be at the expense of being limited to using a subject-dependent training and/or biomechanical data from multiple body locations. The findings could inform a number of applications in the field of healthcare monitoring and developing advanced lower-limb assistive devices by providing insights into classification schemes capable of handling non-steady-state and unstructured locomotion in individuals with mild Parkinson's disease.

## I. INTRODUCTION

Parkinson's disease (PD) is a neurodegenerative disorder of the central nervous system affecting approximately 40 million people worldwide [1]. PD is characterized by a number of motor impairments and gait disorders such as tremor, postural instability, bradykinesia and rigidity [2]. To monitor the progression of the disease and to measure the efficacy of the treatments, accurate tracking of individual's motor activities is essential. Current approaches for evaluating the motor function of individuals with PD are limited to the observer-based and self-reported methods [3]. In observer-based assessment, patients are required to travel to a clinic to perform a set of pre-defined tests. The self-reported approach requires individuals to periodically answer a list of questions about their daily activities. Although useful and currently applied in clinical practice, such evaluations may have some limitations. For instance, they are limited to only a few sessions per year and are costly and inconvenient for both patients and medical providers. They are also subjective and do not adequately reflect motor activities in a free-living environment [4]. Thus, there is a need for developing systems that are convenient and provide quantitative measures of ambulatory performance.

An activity recognition system could provide clinicians with a quantitative profile of motor function behavior in natural settings and over prolonged periods of time, which could further assist them to objectively adapt treatment strategies. Individuals with PD are more susceptible to fall-related injuries due to postural instability and gait disturbances [5], [6]. Real-time monitoring of PD patients' locomotion could provide important information about the risk of falls, which could be used subsequently to apply timely interventions and prevent associated injuries leading to better quality of life [7], [8]. Physical activity monitoring could also complement current approaches for detecting disease-specific predictors such as tremor, bradykinesia or hyperkinesia [9], [10] to distinguish the symptoms during various locomotor activities. Furthermore, the ability to accurately identify individuals intended locomotion could help inform the control of assistive devices [11].

While activity recognition has received significant attention, few studies have applied that specifically to individuals with mobility disorders. In PD, neurological disorders caused by the disease such as altered gait, tremor, and limited mobility have the potential to complicate and adversely affect the monitoring of patient's physical activity. Studies such as [9], [10], [12] have reported on activity monitoring of individuals with PD, however there are limitations that need to be addressed. First, the tasks did not comprehensively represent the activities of daily living, focusing instead on recognition of static/steady-state tasks (e.g., walking, sitting, and standing) performed in isolation. Individuals encounter uneven terrain environments (e.g., stairs and ramps), perform dynamic activities and transition from one task to another in their home and community. Deficits in task switching in individuals with mild Parkinson's disease challenge their ability to unconsciously shift their attention from one task to another [13], [14]. This further emphasizes the significance of developing task recognition frameworks capable of handling unstructured and non-steady-state activities. Second, previous studies primarily relied on input data from entire body, or multiple segments such as trunk, shanks, forearms and thighs [15], [16]. Capturing data from multiple body locations encumbers the patient [17], makes the classification problem more complicated, and increases computation time [18]. An important consideration in activity recognition frameworks is to identify the locations of the body providing the best ability to discriminate between tasks with minimum number of input signals. Furthermore, previous research has primarily focused on within-subject analysis [10], and the generalizability of such studies to subject-independent scenarios has been an unanswered question.

In this study, we collected data from both healthy subjects and PD patients performing a set of unstructured and non-steady state activities. The tasks were designed to challenge cognitive impairment (e.g., difficulty with set shifting) of the individuals with PD [14], [13]. An offline analysis using two commonly used classifiers including a linear discriminant analysis (LDA) and a Long-Short Term Memory (LSTM) neural network was performed for task recognition. The generalizability of two user-independent training paradigms on healthy subjects and PD patient data to a novel subject was tested. Subsequently, the results were compared to subject- dependent training. The performance of accelerographic and gyroscopic data from bilateral foot, forearm, trunk-pelvis, and their fusion were tested within training paradigms. We hypothesized that a more complex classifier (i.e., LSTM) would be more appropriate for modeling non-steady-state tasks, and would outperform LDA. We further hypothesized that collecting data from multiple body locations might not always be necessary depending on the employed classification algorithm and training data.

## II. METHODS

*A. Subjects and Data Collection*

Five healthy subjects (4 males, 1 females, age 25.2±2.5 years, height 1.75±0.11 m, mass 66.8±12.2 kg) and five individuals with early stage PD (2 males, 3 females, age 62.8±3.9 years, height 1.72±0.03 m, mass 77.5±17.88 kg, Hoehn and Yahr stage 1 or 2) participated in the study after providing written informed consent to participate in the protocol approved by the Institutional Review Board at The University of Texas Southwestern Medical Center. Patients did not have a deep brain stimulator implanted. Sixty-six reflective markers were attached to anatomical body locations to track 12 body segments of the arms, legs and torso. A 10-camera optical motion capture system (Vicon, Motion Systems Ltd, UK) was used to capture marker trajectories at 100 Hz in three-dimensional space. Experimental setup consisted of a "terrain park" circuit including an over-ground walkway, a four-step staircase with step height of 0.15 m and depth of 0.30 m, a 2.5 m ramp inclined at 10°, and elevated platforms to connect the stairs and ramp (Fig. 1A). The platform contained a single step of height 0.15 m. Individuals with PD were asked to walk at their comfortable speed and perform five trials of the circuit for both left leading and right leading legs in the following orders: stair ascent/ramp descent and ramp ascent/stair descent. They were instructed to use handrails when desired. Healthy subjects performed five trails of the circuit while using the handrails, and five sets without using the handrails.

*B. Signal Processing and Classification Schemes*

Accelerographic and gyroscopic information of anatomical body segments including feet, trunk-pelvis, and forearms were calculated in three-dimensional trajectories and expressed in local segment coordinate systems using Visual3D (C-Motion, Inc., MD, USA). The tasks included ramp ascent (RA), ramp descent (RD), stair ascent (SA), stair descent (SD), and level-ground walking (LW). In the training data set, changes of direction on the elevated platform, level-walking data that followed stair/ramp, and level walking preceding stair/ramp (Fig. 1A) were all marked as LW. However, level-ground walking data that preceded and followed stair/ramp were tested separately and labeled as LWp and LWf respectively. The beginning of each locomotor mode was marked as the last toe-off of the transitioning leg on the previous terrain. Data were exported to MATLAB (MathWorks, Natick, MA) for further analysis.

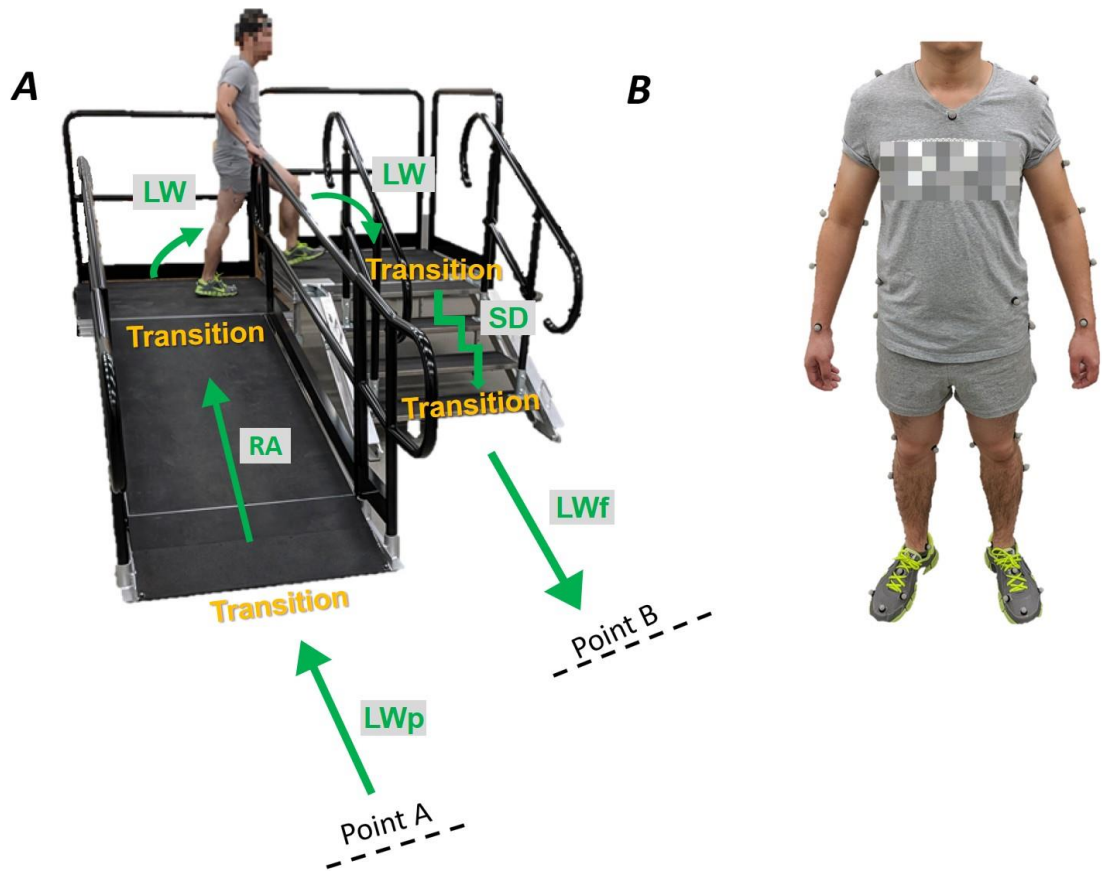

Fig. 1 A) "Terrain park" circuit setup was comprised of a four-step staircase, a ramp and elevated platforms. Subjects performed trials of the circuit in the following orders: They started at point A, performed the locomotion as shown, and stopped at point B. They executed the tasks in the reverse order in the next trial, starting at point B and ending at point A. Circuit trials were performed for both left leading and right leading legs. B) Sixty-six reflective markers were attached to anatomical body locations to track 12 body segments of the arms, legs, and torso.

In order to classify the locomotor activities of individuals with PD, the following classification algorithms, training paradigms, and signal sources were studied.

*Classification algorithms*

- Linear discriminant analysis (LDA)

- Long-short-term memory (LSTM) recurrent neural network. LSTM network parameters were set as follows: batch size=50, number of epochs=70, number of layers=100

*Training paradigms*

- Subject independent I: The classifiers were trained on able-bodied data and evaluated on PD patient's data.

- Subject independent II: The classifiers were trained on PD patients' data, leave-one-subject-out was performed across the patients for model evaluation.

- Subject dependent: Training and testing were performed within trials of each PD patient's data using cross-validation.

*Signal sources*

- Feet
- Trunk-pelvis
- Forearms
- Signal fusion (combination of feet, trunk-pelvis, and forearms data)

Signals were divided into sliding and overlapping analysis windows of size 500 ms with 250 ms increment [19]. The classifiers associated each window to one of the locomotor tasks. To classify the tasks using LDA, six time-domain features including minimum, maximum, mean, standard deviation, first and last sample of each window were extracted [20]. These features are computationally inexpensive and have functioned relatively well in intent recognition frameworks [21], [22]. LSTM was applied on raw data without employing feature extraction. The number of neurons in the input layer was adjusted according to the number of input signals. For instance, for feet, forearms, and trunk-pelvis data, input layer was comprised of 12 neurons as we utilized bilateral accelerographic and gyroscopic signals in three dimensional trajectories. For the combination of all signals (signal fusion) input layer had 36 neurons. Parameters optimized for LSTM include batch size, number of epochs, and number of hidden units. The optimal value for the parameters were selected not just based on the best outcome but also considering the computation time needed to reach that outcome. For instance, while 200 epochs often provided better outcomes relative to 70, the improvement was negligible. Thus, 70 was selected as the optimal number of epochs. Using similar approach, batch size and the number of hidden units were selected as 50 and 100 respectively. A cross-entropy cost function was used to compare the predicted value with the real value during each epoch. Then, Adam optimizer was applied to reduce the loss function values by updating networks weights [23].

*C. System Evaluation*

The subject-independent paradigms were evaluated as follows: in subject independent I, classifiers were trained on only able-bodied data and were tested on all the PD patient data. The results were then averaged across the subjects. In subject independent II, leave-one-subject-out cross validation was performed across PD patients' data. Each time, data from one patient was left out, and the model was trained on the remaining patients' data. The one patient data not included in the training step was used to test the performance of the model. The results were reported as the average across the patients. In subject-dependent paradigm, data from the same patient was used in the training and test sets. Within patient leave-one-trial-out cross validation was performed to evaluate the models. At each evaluation step, one trial of a given task was excluded from the training set and was used to test the model. To evaluate the models, we used F1 score which is the harmonic mean of precision and recall (1). The F1 score is typically employed for imbalanced datasets where some classes have larger number of samples compared to others. In such scenarios using accuracy as an evaluation metric can be misleading since the majority class could be classified with high accuracy while the minority class is highly misclassified.

$$F1\ score = 2 \times \frac{precision \times recall}{precision + recall} \quad (1)$$

In (1), precision is the number of correctly classified samples out of total number of samples classified as the target class. Recall presents correctly classified samples out of the total samples of the target class. Confusion matrices were also computed to quantify the classification results of the proposed scenarios. They provide information about the number of correctly classified as well as misclassified windows across the subjects during each locomotor task. We performed analysis of variance (ANOVA) with the factors being classification algorithms, training paradigms, and signal sources. Post-hoc tests were performed where statistically significant effects were reported ($\alpha=0.05$).

## III. RESULTS

Within each training paradigm, comparing the F1 scores of a given signal source showed improved performance using LSTM compared to LDA, although the differences were not always statistically significant (Table 1). Superior performance of LSTM to LDA was most notable in LWp and when trained on able-bodied data where LDA provided F1 scores ranging 0.19-0.39, while LSTM significantly increased the outcomes to 0.84-0.9. Comparing F1 scores across training paradigms did not demonstrate any statistically significant differences between subject-independent paradigms trained on able-bodied and PD patient's data (p > 0.05). However, improved performance of subject-dependent relative to subject-independent paradigms was observed. The significant improvement (12-133%) of subject-dependent versus subject-independent training was most notable when LDA was applied.

In RA, using subject-independent paradigms with LDA did not result in accurate recognition (F1 score < 0.8) (Table 1). RA was highly misclassified as SA and LW (Table 2). In order for RA to be detected relatively accurate (F1 score ≥ 0.8), using LSTM in subject-independent paradigms appeared to be necessary. Within subject-independent paradigms, signal fusion provided the highest performance (F1 score=0.91) when LSTM was trained on PD patients data, although there was not a statistically significant difference between signal fusion and feet data. Training LSTM on able-bodied data provided F1 scores of 0.85-0.87, and no significant differences across signal sources were observed. RA was best classified using the subject-dependent paradigm and LSTM, where all signal sources provided very accurate outcomes (F1 score= 0.94-0.97).

Similar results were obtained for RD, where using LDA with subject-independent paradigms did not provide accurate detection of the locomotion (F1 score < 0.8) (Table 1). RD was highly confused with SD and LW (Table 2). However, using LSTM and signal fusion in these paradigms improved F1 scores to 0.95. In subject-dependent paradigms, LDA appeared to provide relatively accurate outcomes using all signal sources (F1 score= 0.82-0.9), and LSTM led to a highly accurate recognition (F1 score= 0.96-1). F1 scores of below 0.8 were reported for SA using LDA with subject-independent paradigms except when signal fusion with able-bodied training data were employed (F1 score=0.91) (Table 1). SA was mostly confused with RA and LW (Table 2). However, LSTM led to F1 scores of 0.85-0.92 for feet and signal fusion. In subject-dependent paradigm, both LDA and LSTM appeared to provide improved outcomes (F1 score=0.86-0.98) for all signal sources. In SD, only a few signal sources provided relatively accurate (F1 score ≥ 0.8) recognition of the mode using LDA and subject-independent paradigms (Table 1). SD was highly confused with RD (Table 2). Applying LSTM, however, led to F1 scores of 0.91-0.96 using feet, trunk-pelvis, and signal fusion. SD was best classified when subject-dependent training data was employed (F1 score=0.85-1).

Using LDA with subject-independent paradigms resulted in a very poor recognition of LWp (F1 score=0.19-0.6). Even in subject-dependent training all signal sources except signal fusion demonstrated relatively low F1 scores (0.6-0.66) using LDA. LWp was mostly misclassified as RA/RD (Table 2). However, applying LSTM significantly improved the outcomes to 0.82-0.95. Higher F1 scores were obtained for LWf compared to LWp in most cases, with LSTM outperforming LDA in all training paradigms. Best outcomes were achieved when LSTM was used with subject-dependent training data where F1 scores of 0.98-1 were reported.

## IV. DISCUSSION

Locomotion identification strategies have the potential to be complicated by Parkinsonism associated gait disturbances such as slowed movements, rigidity, tremor, and postural instability which could affect the generalizability of the outcomes obtained in healthy subjects to patient populations. Reduced self-regulating mechanisms could highly challenge patient transitions from one task to another throughout the course of disease [24], [25], [13], [14], and negatively impact the detection of non-steady-state locomotor tasks. Identifying reliable sources of information, appropriate training data, and classification algorithms could significantly improve system outcomes and patient convenience [26]. Therefore, the purpose of this study was to introduce a framework for continuous classification of non-steady-state activities of individuals with PD and investigate the benefits of different classification schemes for an accurate user intent recognition.

Our first hypothesis regarding better performance of LSTM relative to LDA was supported. Within subject-independent paradigms, using a given signal source data with LSTM outperformed LDA in different locomotor tasks. This was especially notable in LWp where LDA resulted in poor task detection (F1 score= 0.19-0.6) while LSTM remarkably improved the outcomes (F1 score=0.76-0.91). When the locomotion involves combinations of non-steady-state activities (e.g., circuit trials in this study), defining the exact boundaries of non-isolated tasks becomes very challenging. In such scenarios, a given task will have biomechanical characteristics of both the previous and the next activity [27] which could negatively impact the performance of task classification approaches. This is especially reflected when the duration of the task is short, so there is not enough time for biomechanical signals to be adjusted to the ongoing task rather than pervious or next mode. For instance, in this study, level ground turns before and after the single step on the elevated platform (Fig. 1A), were marked as level walking during classifier training, while they have dominant biomechanical characteristics of the following and the preceding uneven terrains. Learning such complex patterns in the training data and distinguishing between level walking and other modes would be difficult problems for a linear classifier. High misclassification rates are indicative of the same fact (Table 2). According to our findings, models with non-linear decision boundaries (e.g., LSTM) could be more appropriate for modeling non-steady-state locomotion especially in complex datasets such as subject-independent paradigms.

In subject-independent paradigms, the number of signal sources providing relatively accurate (F1 score ≥ 0.8) recognition appeared to be higher using LSTM compared to LDA (Table 1). For example, using LDA for detection of RA, RD and LWp resulted in F1 scores < 0.8 for all signal sources. Similar results were observed for SA, SD, and LWf where only a few signal sources provided relatively accurate outcomes. However, when LSTM was applied, at least two/three signal sources reached F1 scores of 0.8-0.95. This could suggest higher flexibility in selecting input signal's location using LSTM compared to LDA. The results also support the second hypothesis, highlighting the fact that using a more complex classification algorithm could provide simpler alternatives to collecting data from multiple body locations. From a practical standpoint, this could improve computational complexity, patient convenience, and instrumentation cost by eliminating the need for sensorizing multiple body segments [28], [29]. For instance, feet signals demonstrated comparable performance to signal fusion in all locomotor tasks when LSTM was applied ($p > 0.05$), suggesting feet inertial data as the optimal input information that could properly function across a range of activities with minimal instrumentation.

Statistically significant differences were observed between subject-dependent training relative to subject-independent paradigms in most cases when LDA was applied. The lower accuracy of subject-independent paradigms may be indicative of biomechanical differences between healthy subjects and PD patients [30] as well as across PD patients [31], [32]. This could result in high intra-class variations posing a difficult problem for a linear classifier [33]. However, LSTM appeared to generalize better to such differences. Using LSTM led to achieving more comparable outcomes for subject-independent and subject-dependent paradigms ($p > 0.05$). For instance, comparing feet/signal fusion outcomes using LSTM across training paradigms did not reveal any statistically significant differences between subject-independent and subject-dependent training. This implies that LSTM could allow building subject-independent activity recognition systems. Unlike subject dependent, they are flexible enough to be applied on different users without the need of retraining the model for each person. This would be of higher benefit in individuals with PD where training the system for each user could be inconvenient due to large number of tasks and increased risk of falls, stumbles and injuries during some activities (e.g., non-steady-state transitions).

Neural networks (e.g., LSTM) can receive raw data with minimum pre-processing, alleviating the need for manual feature engineering, thus could minimize engineering bias. Frequency or time-domain features [20], [34] used in conventional machine learning algorithms (e.g., LDA) are problem-specific, and do not generalize well to other problems. For instance, the optimum feature set could vary depending on the target activity. However, conventional algorithms are usually mathematically simple and computationally inexpensive, and do not require large amount of training data. Nonetheless, in this study, employing a mathematically simple classification algorithm such as LDA was shown to be at the expense of being limited to using a training paradigms with lower variability (e.g., subject-dependent) and/or instrumenting multiple body locations. Continuous task classification implemented in this study has the capability to classify data as they are being captured, which is crucial in the context of developing task monitoring scenarios and assistive technologies. In individuals with robotic orthosis/exoskeleton, it would increase the intuitiveness/volitional behavior of the device and enables smooth transitions between locomotor activities. Continuous classification would also allow adaptive assistance [35], predicting fall risks and intervening on these risks to mitigate falls [36]. The reduced levels of flexibility to adapt to new tasks and difficulty in performing transitions in individuals with PD [24], [25] further highlights the advantages of developing such frameworks for tracking characteristics of transitional periods as well as to track steady-state progress.

The study has some limitations. We considered the toe-off event during the transition period as the initiation of the upcoming task. Toe-off could be a relatively accurate approximation of the task initiation where it occurs close to the physical transition point. However, toe-offs occurred at greater distances may negatively impact the outcomes, since a large portion of the gait cycle labeled as the upcoming locomotor task is still within the previous mode. This could result in high misclassification rates especially during the transition period. In future studies, the problem could be addressed by either modifying task separation events, or separating data into steady-state and transitional periods, and performing separate evaluation for each state [37]. Another limitation of this study is that we used motion capture data and not the data from wearable IMU sensors. Inherent errors in actual IMUs such as bias and drift [38] may affect system performance, thus should be taken into consideration in future studies. Further, a small sample of subjects with mild levels of PD participated in this study. Disease associated symptoms such as tremor and bradykinesia could be more severe in patients with higher stages of the PD, which may deteriorate the performance of classification algorithms. Future research should investigate the effects of using data from varied levels of disease severity and a larger subject pool to accommodate for the potential across-subject variabilities.

## V. CONCLUSION

We introduced a task recognition framework for tracking relatively unstructured locomotor activities in individuals with mild PD. Our results demonstrated that, models with non-linear decision boundaries (e.g., LSTM) could be more appropriate relative to linear classifiers (e.g., LDA) for modeling non-steady-state locomotion. LSTM could provide simpler alternatives (e.g., feet data) to collecting data from multiple locations improving user convenience and system's computational complexity for its eventual clinical use. The model could also allow building subject-independent activity recognition systems that are flexible enough to be applied on different users without the need of retraining the model each time. These findings could provide insights into designing activity recognition frameworks for healthcare monitoring and lower-limb assistive devices improving system efficacy and user convenience without sacrificing accuracy.

TABLE I

F1 SCORES FOR CLASSIFICATION OF TARGET LOCOMOTOR TASKS USING SIGNALS FROM VARYING BODY SEGMENTS AND ACROSS DIFFERENT CLASSIFICATION ALGORITHMS AND TRAINING PARADIGMS.

| | | Signal source | RA | RD | SA | SD | LWp | LWf |
|---|---|---|---|---|---|---|---|---|
| **Subject Independent I** | LDA | Feet | 0.56 (0.07) | 0.69 (0.08) | 0.79 (0.08) | 0.81 (0.1) | 0.39 (0.23) | 0.62 (0.12) |
| | | Trunk-pelvis | 0.67 (0.05) | 0.62 (0.15) | 0.79 (0.08) | 0.60 (0.33) | 0.19 (0.22) | 0.89 (0.03) |
| | | Forearms | 0.44 (0.08) | 0.42 (0.22) | 0.57 (0.13) | 0.41 (0.15) | 0.32 (0.19) | 0.57 (0.09) |
| | | Fusion | 0.63 (0.07) | 0.78 (0.1) | 0.91 (0.08) | 0.90 (0.06) | 0.39 (0.25) | 0.80 (0.07) |
| | LSTM | Feet | 0.87 (0.1) | 0.94 (0.04) | 0.92 (0.04) | 0.96 (0.02) | 0.90 (0.04) | 0.92 (0.05) |
| | | Trunk-pelvis | 0.87 (0.18) | 0.95 (0.05) | 0.62 (0.35) | 0.94 (0.07) | 0.84 (0.21) | 0.92 (0.03) |
| | | Forearms | 0.85 (0.1) | 0.68 (0.2) | 0.84 (0.1) | 0.74 (0.17) | 0.84 (0.09) | 0.82 (0.1) |
| | | Fusion | 0.86 (0.11) | 0.95 (0.03) | 0.92 (0.01) | 0.95 (0.03) | 0.90 (0.06) | 0.92 (0.05) |
| **Subject Independent II** | LDA | Feet | 0.61 (0.08) | 0.70 (0.04) | 0.79 (0.12) | 0.85 (0.04) | 0.55 (0.22) | 0.59 (0.22) |
| | | Trunk-pelvis | 0.68 (0.1) | 0.64 (0.17) | 0.52 (0.32) | 0.60 (0.27) | 0.59 (0.27) | 0.74 (0.15) |
| | | Forearms | 0.46 (0.21) | 0.50 (0.17) | 0.40 (0.18) | 0.64 (0.04) | 0.58 (0.19) | 0.59 (0.2) |
| | | Fusion | 0.78 (0.11) | 0.70 (0.12) | 0.74 (0.25) | 0.68 (0.35) | 0.60 (0.37) | 0.74 (0.16) |
| | LSTM | Feet | 0.89 (0.07) | 0.85 (0.14) | 0.86 (0.08) | 0.92 (0.05) | 0.83 (0.12) | 0.89 (0.07) |
| | | Trunk-pelvis | 0.77 (0.31) | 0.89 (0.16) | 0.73 (0.34) | 0.90 (0.05) | 0.77 (0.33) | 0.88 (0.08) |
| | | Forearms | 0.73 (0.21) | 0.76 (0.19) | 0.67 (0.22) | 0.64 (0.18) | 0.76 (0.1) | 0.77 (0.13) |
| | | Fusion | 0.91 (0.07) | 0.95 (0.03) | 0.85 (0.08) | 0.91 (0.1) | 0.91 (0.05) | 0.91 (0.06) |
| **Subject Dependent** | LDA | Feet | 0.78 (0.09) | 0.81 (0.07) | 0.94 (0.05) | 0.91 (0.04) | 0.65 (0.17) | 0.78 (0.09) |
| | | Trunk-pelvis | 0.84 (0.09) | 0.82 (0.06) | 0.88 (0.06) | 0.90 (0.04) | 0.66 (0.17) | 0.87 (0.04) |
| | | Forearms | 0.79 (0.08) | 0.82 (0.06) | 0.86 (0.08) | 0.85 (0.05) | 0.60 (0.13) | 0.83 (0.05) |
| | | Fusion | 0.92 (0.06) | 0.90 (0.06) | 0.95 (0.06) | 0.93 (0.04) | 0.91 (0.05) | 0.92 (0.04) |
| | LSTM | Feet | 0.94 (0.06) | 1.00 (0.01) | 0.98 (0.03) | 1.00 (0.01) | 0.90 (0.1) | 1.00 (0.0) |
| | | Trunk-pelvis | 0.98 (0.03) | 0.96 (0.05) | 0.94 (0.08) | 0.98 (0.02) | 0.92 (0.11) | 0.98 (0.02) |
| | | Forearms | 0.97 (0.03) | 0.97 (0.03) | 0.95 (0.05) | 0.99 (0.02) | 0.82 (0.25) | 0.99 (0.01) |
| | | Fusion | 0.99 (0.02) | 0.99 (0.01) | 0.98 (0.03) | 0.99 (0.01) | 0.95 (0.07) | 0.99 (0.01) |

TABLE II

CONFUSION MATRICES FOR CLASSIFICATION OF TARGET LOCOMOTOR TASKS USING SIGNALS FROM VARYING BODY SEGMENTS AND ACROSS DIFFERENT CLASSIFICATION ALGORITHMS AND TRAINING PARADIGMS. DIAGONAL ELEMENTS SHOW THE NUMBER OF CORRECTLY CLASSIFIED WINDOWS FOR THE TARGET TRANSITIONS, AND OFF-DIAGONAL ELEMENTS ARE INDICATIVE OF MISCLASSIFICATION.

| | | | *Subject Independent I* | | | | | *Subject Independent II* | | | | | *Subject Dependent* | | | | |
|---|---|---|---|---|---|---|---|---|---|---|---|---|---|---|---|---|---|
| | | | RA | RD | SA | SD | LW | RA | RD | SA | SD | LW | RA | RD | SA | SD | LW |
| *Feet* | LDA | RA | **284** | 0 | 73 | 0 | 16 | **301** | 0 | 26 | 0 | 46 | **365** | 0 | 0 | 0 | 8 |
| | | RD | 10 | **289** | 0 | 68 | 40 | 15 | **333** | 0 | 40 | 19 | 3 | **391** | 2 | 4 | 7 |
| | | SA | 24 | 0 | **308** | 0 | 1 | 45 | 0 | **269** | 1 | 18 | 14 | 1 | **316** | 0 | 2 |
| | | SD | 0 | 9 | 0 | **275** | 0 | 0 | 11 | 0 | **272** | 1 | 0 | 6 | 0 | **275** | 3 |
| | | LWp | 69 | 92 | 12 | 4 | **88** | 29 | 100 | 4 | 3 | **129** | 57 | 73 | 11 | 1 | **123** |
| | | LWf | 261 | 40 | 49 | 60 | **408** | 210 | 105 | 33 | 45 | **425** | 135 | 93 | 12 | 44 | **534** |
| | LSTM | RA | **301** | 0 | 13 | 0 | 59 | **316** | 0 | 0 | 0 | 57 | **368** | 0 | 5 | 0 | 0 |
| | | RD | 0 | **373** | 0 | 6 | 28 | 0 | **340** | 1 | 3 | 63 | 0 | **404** | 0 | 2 | 1 |
| | | SA | 5 | 0 | **300** | 0 | 28 | 7 | 0 | **266** | 0 | 60 | 2 | 0 | **330** | 0 | 1 |
| | | SD | 0 | 0 | 0 | **273** | 11 | 0 | 1 | 1 | **256** | 26 | 0 | 0 | 0 | **284** | 0 |
| | | LWp | 3 | 4 | 14 | 0 | **244** | 13 | 18 | 1 | 1 | **232** | 36 | 1 | 1 | 0 | **227** |
| | | LWf | 11 | 4 | 0 | 10 | **793** | 1 | 6 | 4 | 4 | **803** | 0 | 0 | 0 | 0 | **818** |
| *Trunk-pelvis* | LDA | RA | **319** | 0 | 41 | 0 | 13 | **307** | 0 | 21 | 0 | 45 | **352** | 0 | 7 | 0 | 14 |
| | | RD | 5 | **277** | 0 | 90 | 35 | 6 | **301** | 0 | 65 | 35 | 8 | **357** | 0 | 15 | 27 |
| | | SA | 75 | 1 | **255** | 0 | 2 | 103 | 1 | **158** | 0 | 71 | 18 | 1 | **312** | 0 | 2 |
| | | SD | 0 | 108 | 0 | **176** | 0 | 0 | 104 | 0 | **180** | 0 | 0 | 16 | 0 | **267** | 1 |
| | | LWp | 152 | 55 | 10 | 6 | **42** | 16 | 55 | 28 | 5 | **161** | 34 | 29 | 65 | 2 | **135** |
| | | LWf | 27 | 64 | 4 | 18 | **705** | 87 | 90 | 20 | 43 | **578** | 57 | 69 | 0 | 31 | **661** |
| | LSTM | RA | **349** | 0 | 0 | 0 | 24 | **272** | 2 | 40 | 0 | 59 | **364** | 3 | 0 | 0 | 6 |
| | | RD | 0 | **380** | 0 | 2 | 25 | 0 | **382** | 1 | 2 | 22 | 0 | **398** | 0 | 2 | 7 |
| | | SA | 102 | 0 | **153** | 0 | 78 | 81 | 0 | **194** | 0 | 58 | 0 | 0 | **316** | 0 | 17 |
| | | SD | 0 | 8 | 0 | **272** | 4 | 0 | 64 | 0 | **203** | 17 | 0 | 0 | 0 | **275** | 9 |
| | | LWp | 25 | 4 | 0 | 0 | **236** | 26 | 9 | 6 | 0 | **224** | 7 | 15 | 26 | 0 | **217** |
| | | LWf | 1 | 7 | 0 | 27 | **783** | 1 | 37 | 10 | 17 | **753** | 0 | 0 | 0 | 1 | **817** |
| *Forearms* | LDA | RA | **228** | 4 | 27 | 0 | 114 | **228** | 0 | 77 | 0 | 68 | **344** | 0 | 12 | 0 | 17 |
| | | RD | 1 | **195** | 0 | 145 | 66 | 17 | **218** | 2 | 113 | 57 | 7 | **373** | 0 | 15 | 12 |
| | | SA | 161 | 0 | **166** | 0 | 6 | 143 | 1 | **158** | 0 | 31 | 21 | 2 | **307** | 0 | 3 |
| | | SD | 0 | 99 | 0 | **174** | 11 | 0 | 55 | 0 | **225** | 4 | 0 | 16 | 0 | **264** | 4 |
| | | LWp | 96 | 100 | 2 | 10 | **57** | 14 | 62 | 20 | 16 | **153** | 42 | 59 | 56 | 1 | **107** |
| | | LWf | 173 | 6 | 31 | 203 | **405** | 129 | 71 | 126 | 57 | **435** | 80 | 59 | 8 | 59 | **612** |
| | LSTM | RA | **327** | 0 | 14 | 0 | 32 | **258** | 16 | 12 | 0 | 87 | **369** | 0 | 1 | 0 | 3 |
| | | RD | 0 | **212** | 0 | 56 | 139 | 4 | **317** | 0 | 10 | 76 | 6 | **397** | 0 | 2 | 2 |
| | | SA | 7 | 0 | **265** | 0 | 61 | 53 | 0 | **203** | 1 | 76 | 2 | 0 | **325** | 0 | 6 |
| | | SD | 0 | 0 | 0 | **211** | 73 | 0 | 57 | 0 | **173** | 54 | 0 | 0 | 0 | **277** | 7 |
| | | LWp | 6 | 4 | 11 | 0 | **244** | 12 | 9 | 35 | 1 | **208** | 14 | 6 | 22 | 0 | **223** |
| | | LWf | 62 | 1 | 4 | 14 | **737** | 32 | 46 | 4 | 84 | **652** | 0 | 2 | 0 | 0 | **816** |
| *Fusion* | LDA | RA | **324** | 0 | 25 | 0 | 24 | **330** | 0 | 5 | 0 | 38 | **370** | 0 | 0 | 0 | 3 |
| | | RD | 28 | **277** | 0 | 25 | 77 | 6 | **332** | 0 | 19 | 50 | 2 | **378** | 1 | 7 | 19 |
| | | SA | 12 | 0 | **316** | 0 | 5 | 46 | 1 | **232** | 0 | 54 | 10 | 0 | **317** | 0 | 6 |
| | | SD | 0 | 2 | 0 | **281** | 1 | 0 | 73 | 0 | **206** | 5 | 0 | 5 | 0 | **277** | 2 |
| | | LWp | 150 | 14 | 8 | 1 | **92** | 24 | 54 | 7 | 1 | **179** | 21 | 10 | 6 | 1 | **227** |
| | | LWf | 139 | 11 | 10 | 38 | **620** | 89 | 91 | 12 | 40 | **586** | 28 | 40 | 8 | 33 | **709** |
| | LSTM | RA | **304** | 0 | 6 | 0 | 63 | **341** | 0 | 1 | 0 | 31 | **373** | 0 | 0 | 0 | 0 |
| | | RD | 0 | **389** | 0 | 2 | 16 | 0 | **384** | 0 | 2 | 21 | 0 | **403** | 0 | 2 | 2 |
| | | SA | 6 | 0 | **298** | 0 | 29 | 23 | 0 | **254** | 0 | 56 | 0 | 0 | **330** | 0 | 3 |
| | | SD | 0 | 0 | 0 | **274** | 10 | 0 | 0 | 0 | **262** | 22 | 0 | 0 | 0 | **284** | 0 |
| | | LWp | 11 | 4 | 5 | 0 | **245** | 4 | 5 | 3 | 0 | **253** | 11 | 12 | 8 | 0 | **234** |
| | | LWf | 1 | 15 | 2 | 15 | **785** | 9 | 8 | 4 | 17 | **780** | 1 | 1 | 0 | 0 | **816** |